\documentstyle[preprint,aps]{revtex}
\begin{document}

\newcommand{\btu}{\bigtriangledown}
\newcommand{\bge}{\begin{equation}}
\newcommand{\ege}{\end{equation}}
\newcommand{\bga}{\begin{eqnarray}} 
\newcommand{\ega}{\end{eqnarray}}
\newcommand{\nnu}{\nonumber}

\draft
\preprint{\vbox{\hbox{IP/BBSR/95-66}\hbox{hep-th/9507063}}} 

\title {\bf S-DUALITY AND COSMOLOGICAL CONSTANT IN STRING THEORY }
\author{\sc Supriya Kar$^{1}$, Jnanadeva Maharana$^{2}$
\footnote{Jawaharlal Nehru Fellow } 
and Harvendra Singh$^{2}$
\footnote{e-mail: maharana, hsingh@iopb.ernet.in}} 
\address{ $^1$ Mehta Research Institute of Mathematics \&
Mathematical Physics, \\10, Kasturba Gandhi Marg, Allahabad-211
002, India }
\address{  $^2$ Institute of Physics, Bhubaneswar 751 005, India}
\maketitle

\begin{abstract} 
The S-duality invariance of the equations of motion of four
dimensional string effective action with cosmological constant,
$\Lambda $, is studied. It is demonstrated that the S-duality
symmetry of the field equations are broken for nonzero
$\Lambda$. The ``naturalness'' hypothesis is invoked to argue that
$\Lambda $ should remain small since exact S-duality symmetry
will force the cosmological constant to vanish in the string effective
action. 
\end{abstract}
\vspace{1cm}
\begin{center}
( To appear in Phys. Lett. B )
\end{center}

\narrowtext 

\newpage

\par It is recognized that S-duality is an important symmetry of
string theory which relates the strong and weak coupling
domains. The consequences of this symmetry \cite{1} are
interesting and surprising. In the recent past, a variety of 
novel results have been derived for supersymmetric gauge
theories\cite{2} in sequel to these new developments in string
theory.  

\par The purpose of this article is to explore further the
consequences of S-duality. It is well known that the equations
of motion derived from the effective action are invariant under
S-duality, although the action is not. However, these results
are derived in the absence of the cosmological constant term in the
action. We show that in presence of the cosmological constant,
$\Lambda$, the equations of motion associated with a four
dimensional effective action, obtained through dimensional
reduction, are not invariant under S-duality transformations.
Nevertheless, the invariance of the equations of motion is
recovered once we set $\Lambda=0$. This leads us  to
conjecture that exact S-duality symmetry will force the
cosmological constant to vanish. At this stage, we are tempted
to invoke the hypothesis of naturalness \cite{3} expounded by 't
Hooft which says that a parameter in any theory remains small,
if the symmetry is enhanced by setting that parameter to zero.  
For example, in electrodynamics setting the electron mass $m_e\,=\,0$
enhances the symmetry of the action and the  chiral
symmetry is restored. Therefore, it is 
guaranteed that $m_e$ remains small and the corrections are
proportional to $m_e$ itself. We recall that according to 't
Hooft, the vanishingly  small value of cosmological constant is
unnatural \cite{3}; putting it equal to zero does not seem to
increase the symmetry of Einstein-Hilbert action. Consequently, he
concluded that gravitational effects do not obey naturalness in
conventional theory. However in the framework of string theory, 
$\Lambda $ obeys the naturalness criterion, $i.e.$ putting
$\Lambda =0$ enhances the stringy symmetry. We are aware that
S-duality is not a symmetry of the action and hence may not
satisfy the strict criterion of naturalness. 

The cosmological constant is a parameter measured very close to
zero and it is a vintage theoretical problem to explain the
smallness of $\Lambda $ which has eluded physicists for a long
time \cite{4}. Several attempts have already been
made to explain the vanishing cosmological constant in the framework
of string theory \cite{5}. Recently, Witten \cite{6} has argued
that the vanishing of cosmological constant and the 
absence of massless dilaton might be explained by a duality
between supersymmetric string vacuum in three dimensions and a
non-supersymmetric string vacuum in four-dimensions.
The issue has also been addressed in a more concrete
model by Becker, Becker and Strominger\cite{7}. 

In this investigation, we study four-dimensional string
effective action in the presence of cosmological constant term.
First, we consider an effective action which is derived by
toroidal compactification of a higher dimensional action.
The massless field contents of the resulting four dimensional
superstring action in the bosonic sector are: graviton,
antisymmetric tensor, dilaton, gauge fields and the moduli. In
four dimensions, the antisymmetric 
tensor field can be traded for a pseudoscalar axion $\chi$.
The dilaton, axion and gauge fields transform nontrivially under
S-duality whereas the metric and moduli remain invariant.
The Einstein and matter field equations are derived and
it is shown that these equations are not invariant under S-duality
transformations. Next, we consider an effective action which
admits  charged black hole solutions. It is well known
that, near the horizon, the underlying conformal field theory is
described by tensor product of two WZW models: one corresponds
to an $SL(2,R)$ whereas the other is identified with an $SU(2)$
with WZW lavel constants $K_{SL}$ and $K_{SU}$ respectively. The
corresponding four dimensional string effective action is
recognized as an action obtained from the compactification of a
six dimensional one with a cosmological constant term. We
present yet another way to obtain the four dimensional black
hole solution by axial gauging of the $SL(2,R)$ WZW Lagrangian.
The four dimensional effective action is derived and explicit 
solutions of the background field equations are given.

Now, consider a string effective action in $ D$ spacetime
dimensions with massless fields such as graviton, ${\hat G}_{M N}$,
antisymmetric tensor, ${\hat B}_{M N}$, $( M, N \,=\,
0,1,\cdots,D-1)$, dilaton, ${\hat\Phi}$ and $n$
Abelian gauge fields, ${\hat A}^I_M$. If we compactify coordinates on a
$d\,=\,D-4$ dimensional torus and assume that the backgrounds are
independent of these $d$ compact coordinates, the resulting four
dimensional reduced effective action takes the following form\cite{8},
\newpage
\bga
S = \int d^4 x \sqrt{-g}\,\,e^{-\Phi} \big ( R +
g^{\mu\nu}\partial_\mu\Phi\partial_\nu\Phi 
&&+ {1\over8} Tr \partial_\mu\, M^{-1}\partial^\mu M \nnu\\
&&- {1\over4} {\cal F}^i_{\mu\nu}\, ( M^{-1} )_{i\,j}\,{\cal
F}^{j\,\mu\nu} -{1\over12} H_{\mu\nu\lambda}\,H^{\mu\nu\lambda}
- 2\Lambda \big ), 
\label{13}
\ega
where
\bga
&& {\hat G}_{M N}=\pmatrix{ g_{\mu\nu} + A^{(1)}_{\mu\,\alpha}
   A^{(1)\,\alpha}_\nu & A^{(1)}_{\mu\,\beta}\cr
   A^{(1)}_{\nu\,\alpha} & G_{\alpha\beta}}\ ,\nnu\\ \nnu\\
  && M = \pmatrix{ G^{-1} & -G^{-1} C & -G^{-1} A^T \cr 
-C^T G^{-1} & G+C^T G^{-1} C + A^T A & C^T G^{-1}A^T + A^T \cr
-A G^{-1} & A G^{-1}C +A & 1+ A G^{-1} A^T},
\label{14}
\ega
$C_{\alpha\beta}= {1\over2} A^I_\alpha A^I_\beta +
B_{\alpha\beta}$ and $\Phi = {\hat \Phi} - {1\over2}\ln\det G_{\alpha\beta}$ is
shifted dilaton with the spacetime dependent background fields
$( G_{\alpha\beta}\,, \,\, A^I_\alpha \equiv{\hat
A}^I_\alpha\,,\,\,B_{\alpha\beta}\equiv {\hat B}_{\alpha\beta}
)$ defining a generic point in moduli-space in the toroidal
compactification of  string theory. The moduli
M is a $(2d+n)\times(2d+n)$ matrix valued scalar field and
satisfies the condition $ M {\cal L} M {\cal L} =1$,
 where ${\cal L}$ is the $O(d,d+n)$ metric, 

\begin{eqnarray}
{\cal L} = \pmatrix{ 0 & I_d & 0 \cr I_d & 0 & 0\cr 0 & 0
&I_n}\> ,\>\>\>\>\>\>
\Omega^T {\cal L} \,\Omega = {\cal L}\,.
\label{15}
\end{eqnarray}
Here $I_d$ is d-dimensional identity matrix and $\Omega$ is an element of
the group $O(d,d+n)$. The field strengths appearing in (\ref{13}) are

\bga
H_{\mu\nu\lambda}&=& \partial_\mu B_{\nu\lambda} -{1\over2}{\cal
A}_\mu^i {\cal L}_{i\,j} {\cal F}^j_{\nu\lambda} + cyclic\,\, perm.\nnu\\
{\cal F}^i_{\mu\nu}&=& \partial_\mu {\cal A}^i_\nu -
\partial_\nu{\cal A}^i_\mu \>\> ,\nnu
\ega
where i, j are $O(d,d+n)$ matrix indices. ${\cal A}^i_\mu \equiv 
( \,A^{(1)\alpha}_\mu = {\hat G}_{\mu\alpha}\,,
A^{(2)}_{\mu\alpha} = {\hat B}_{\mu\alpha} + 
{\hat B}_{\alpha\beta} A^{(1)\beta}_\mu
+{1\over2} {\hat A}^I_\alpha \, A^{(3)\,I}_\mu \,,
A^{(3)\,I}_\mu = {\hat A}^I_\mu - {\hat A}^I_\alpha\,
A^{(1)\,\alpha}_\mu\,)$ 
is a $(2d +n )$ component vector field.
It is more convenient for the implementation of S-duality
transformation to rescale
the $\sigma$-model metric to Einstein metric, $g_{\mu\nu}\to
e^{\Phi}g_{\mu\nu}$, and introduce the 
axion $ \partial_\sigma \chi =  {(\eta^2/ 6)} 
\sqrt{-g}\epsilon_{\mu\nu\lambda\sigma} 
H^{\mu\nu\lambda}$ where $\,\eta = e^{-\Phi}$ . Then (1) can be reexpressed as 

\bga
S = \int d^4 x \sqrt{- g} \big ( R -{1\over 2\, \eta^2}
g^{\mu\nu}\partial_\mu\Psi\,\partial_\nu {\bar \Psi } 
&&+ {1\over8} Tr (\partial_\mu M^{-1}\partial^\mu M ) \nnu\\
&&-{1\over4} \eta\, {\cal F}^i_{\mu\nu} M^{-1}_{i\,j}{\cal
F}^{j\,\mu\nu} + {1\over4} \chi \,{\cal F}^i_{\mu\nu}\,{\cal
L}_{i j}\, {\tilde {\cal F}}^{j\,\mu\nu} 
- {2\Lambda\over \eta} \big )\ ,
\label{16}
\ega
where $\Psi = \chi + i \,\eta$ is a  complex scalar field and
\bga
{\tilde{\cal F}}^i_{\mu\nu}&&={1\over2}\sqrt{-g}
\epsilon_{\mu\nu\rho\sigma}\, 
{\cal F}^{i\,\rho\sigma}.
\label{17}
\ega
 We mention in passing that the action (\ref{13}) is manifestly invariant
under global $O(d,d+n)$ transformations:
 \bga
M&\to& \Omega \, M \, \Omega^T\;\; ,\;\;\;\;\; \Omega\>\in
\>O(d,d+n)\nnu\\ \Phi&\to& \Phi,\,\,\, g_{\mu\nu}\to
g_{\mu\nu},\,\,\,B_{\mu\nu}\to B_{\mu\nu}, 
\>\>\> {\cal A}^i_\mu\to \Omega^i_j \,{\cal A}^j_\mu.
 \label{18}
 \ega
The equations of motion corresponding to $\Psi$, $g_{\mu\nu}$
and $A_\mu$ derived from the action (\ref{16}) are

\bge
{\btu_\mu\btu^\mu\Psi\over \eta^2} + i {\btu_\mu\Psi \btu^\mu\Psi\over
\eta^3} - {i\over4}\,{\cal F}\, M \,{\cal F} 
+{1\over4}\,{\cal F}\,{\cal L}\,{\tilde{\cal F}}
+i\, {2 \,\Lambda\over \eta^2}=0 ,
\ege
\bge
R_{\mu\nu} -{\btu_\mu\Psi \btu_{\nu}{\bar \Psi}\over 2\,\eta^2}
+{1\over8} Tr (\partial_\mu M^{-1}\,\partial_\nu M) 
-{\eta\over2} {\cal F}_{\mu\lambda} M^{-1} {\cal F}_\nu^{\,\,\lambda}
+g_{\mu\nu}\left( {\eta\over8} {\cal F}\, M^{-1}\,{\cal F} -
{\Lambda\over\eta} \right) = 0 ,
\ege
\bge
\btu_\mu \left( \eta \,(M\,{\cal L})_{i j}{\cal F}^{j\,\mu\nu} -\chi
\,{\tilde{\cal F}} ^{i\,\mu\nu} \right)=0 ,
\ege
and the Bianchi identity is 
\bge
\btu_\mu {\tilde{\cal F}}^{i\,\mu\nu} = 0 .
\label{19}
\ege
The S-duality transformations correspond to 
\bga
&&\Psi \to {a\,\Psi + b \over c\,\Psi +
d}\,, \>\>\> a\,d - b\,c =1 \;\; , \,\,\,\,\,a,b,\cdots\in \,\,{\bf Z},
\nnu\\ &&{\cal F}^i_{\mu\nu} \to c \,\eta \,( M\,{\cal L})_{i j} 
\,{\tilde {\cal F}}^j_{\mu\nu} +( c \, \chi + d)\,{\cal
F}^i_{\mu\nu} 
\label{20}
\ega
and the metric $g_{\mu\nu}$ and moduli $M$ remain invariant.

Explicit calculations show that under S-duality the terms in
eqs.(7) and (8) remain invariant when $\Lambda = 0$; however for
nonvanishing $\Lambda$ these equations are not S-duality
invariant. In this context, we mention
that it has been observed \cite{ms}, in specific cases, that
S-duality invariance of equations of motion is broken in
presence of $\Lambda $. To analize S-duality invariance
of eqs. (7) and (8), let us consider a specific transformation ( $a=d=0,
 b=-c=1$ )
\begin{equation}
\Psi\to -1/\Psi \,\,{\rm and}\,\,\,\,{\cal F}^i_{\mu\nu} \to - \,\eta \,(
M\,{\cal L})_{i j} \,{\tilde {\cal F}}^j_{\mu\nu} -  \chi \,{\cal
F}^i_{\mu\nu}.
\label{20a}
\end{equation}
Now it is straightforward to find
that first four terms on the left-hand-side of (7) are invariant
under above transformation (\ref{20a}) while the last term with
$\Lambda$ is not. Similarly, it can also be checked  that except
$\Lambda$-term all other terms in eq.(8) make an invariant combination.
Thus in general, the presence of cosmological constant breaks the S-duality
invariance of the  string equations of motion.
Furthermore, in principle the cosmological constant $\Lambda$ can be
generalised to a nontrivial  dilaton potential
$V(\Phi)$ which might be generated due to nonperturbative effects.
However, the corresponding equations of motion are
S-duality invariant only if $V(\Phi)=0$.  
 We write the equations of motion  involving $V(\Phi)$
 ( after rescaling to Einstein metric ):

\begin{eqnarray}
{\bigtriangledown_\mu\bigtriangledown^\mu\Psi\over \eta^2} + i
{\bigtriangledown_\mu\Psi \bigtriangledown^\mu\Psi\over 
\eta^3} - {i\over4}\,{\cal F}\, M \,{\cal F} 
+{1\over4}\,{\cal F}\,{\cal L}\,{\tilde{\cal F}}
+i\,\left( {2 \,\tilde{V}(\eta)\over \eta^2} - {2\over \eta}
{\partial \tilde{V}(\eta)\over\partial\eta}\right)=0 ,\nonumber\\ \nonumber\\
R_{\mu\nu} -{\bigtriangledown_\mu\Psi \bigtriangledown_{\nu}{\bar
\Psi}\over 2\,\eta^2} 
+{1\over8} Tr (\partial_\mu M^{-1}\,\partial_\nu M) 
-{\eta\over2} {\cal F}_{\mu\lambda} M^{-1} {\cal F}_\nu^{\,\,\lambda}
+g_{\mu\nu}\left( {\eta\over8} {\cal F}\, M^{-1}\,{\cal F} -
{\tilde{V}(\eta)\over\eta} \right) = 0 ,
\label{20b}
\end{eqnarray}
where $\tilde{V}(\eta)$ is the dilaton potential reexpressed in
terms of new variable $\eta=e^{-\Phi}$.
We note that the above equations of motion (\ref{20b}) are not
invariant under the transformation (\ref{20a}) as long as the dilaton
potential $V(\Phi)$ is nonzero. 

Now let us consider an example\cite{9} of six dimensional target
space constructed by taking a tensor product of two WZW theories
with the groups $SL(2,R)$ and $SU(2)$ respectively. Thus the
underlying conformal field theory describing the above target
space is exact. If we compactify one coordinate, say $\varphi$,
of $SL(2,R)$ and another $\zeta$ of 
$SU(2)$ on torii then the resulting theory has a metric with
$(- + + +)$ Minkowski signature. A pair of gauge fields $
{\cal A}^{(1)\,\alpha}_\mu$ ($\,\alpha = \varphi, \zeta\,$ )  appear
from the metrics of the two groups and another pair of gauge
fields ${\cal A}^{(2)}_{\mu,\,\alpha}\,$ come from the
antisymmetric tensor fields. The scalar multiplet consists of a
pair from the moduli and the dilaton. The exact
conformal field theory backgrounds that satisfy equations of
motion of four dimensional string effective action (\ref{16}) are

\bga
ds^2 &=& - ( {r^2\over K_{SL}} - M +{J^2\over 4\,r^2}) dt^2 +
( {r^2\over K_{SL}} - M +{J^2\over 4\,r^2})^{-1} dr^2 +
{K_{SU}\over4}(\,d\theta^2 + \sin^2\theta\,d\phi^2\,) \nnu\\ \nnu\\
{\cal A}^{(1)\,\varphi}_\mu &=&\pmatrix {-{J^2\over 2
r^2},&0,&0,& 0},\>\>\>\> {\cal A}^{(1)\,\zeta}_\mu =\pmatrix {0,
&0,&0,& m\,\cos\theta},\nnu\\  \nnu\\
{\cal A}^{(2)}_{\mu,\,\varphi}&=&\pmatrix {-{r^2\over l},&0,&0,&
0},\>\>\>\> {\cal A}^{(2)}_{\mu,\,\zeta}=\pmatrix { 0,&0,& 0, &
\pm {n\over 4} \,\cos\theta },\>\>\>\>\> {\cal
A}^{(3)\ I}_{\mu}=0, \nnu\\ \nnu\\
G_{\alpha\beta}&=&\pmatrix{r^2&0\cr 0&
{n\over4\,m}},\>\>\>B_{\alpha\beta}=0,\>\>\>{\Phi}= - \ln r
+ const. ,\>\>\> H_{\mu\nu\sigma}=0, 
\label{22}
\ega
where $\alpha,\> \beta$ run over two compactified directions
$\varphi$ and $\zeta$. Here $G_{\varphi \varphi}\,=\, r^2$ and
$G_{\zeta \zeta}\,=\, {n/ 4m}$ correspond to the moduli.
The corresponding 
six dimensional string effective action has graviton,
antisymmetric tensor and dilaton only. 

We mention in passing that the compactification of the direction
$\zeta$ gives rise to a $U(1)$ gauge field with magnetic charge
$m$ and modular invariance imposes the constraint
$m\,n\,=\,K_{SU}$.   
Notice that large $K_{SU}$ and $K_{SL}$ limits corresponds to
Bertotti-Robinson \cite{br} space
time in four dimensions. This solution
describes the throat limit of extremal dilaton black holes with
electric and magnetic charge investigated by Kallosh et.
al.\cite{10} and also describes the throat limit of the
Reissner-Nordstrom black hole. We recall that large $K_{SU}$ and
$K_{SL}$ can be envisaged as the classical limit since these
constants play the role of ${1/ \,\hbar}$ in the WZW theory. In this
limit, the cosmological constant $\Lambda\to0$.

Next, we present another way to obtain the four-dimensional
black hole solutions. In this case, instead of compactifying
the coordinate $\varphi$ of the $SL(2,R)$ alluded to above, we
gauge the $U(1)$ subgroup. The gauged $SL(2,R)$ WZW action can
be written in light cone coordinates ($z,\,{\bar z}$)

\bge
S(U,A)=S(U) + {K_{SL}\over 2\pi}\int d^2z Tr\left [  U^{-1}
\partial_z\, U \,A_{\bar z} + \partial_{\bar z} U U^{-1} A_z + 
U^{-1}\,A_z \,U\,A_{\bar z} + A_z\,A_{\bar z}\right ]\> ,
\label{wz}
\ege
where $U(x)\in SL(2,R) \>\not\vdash\> x $ in the manifold $M$.
Integrating out the gauge fields in (\ref{wz}) by taking care of
the Jacobian in the corresponding path-integral, one gets the
two dimensional target space configuration\cite{11}. As a
result, the effective theory is a five-dimensional one with an
appropriate cosmological constant term. The corresponding four
dimensional effective action will arise from the
compactification of this five-dimensional theory. In this
prescription there are only two gauge fields, namely one comming from
the metric and the other from the antisymmetric tensor field
when $\zeta$ coordinate of $SU(2)$ is compactified. The
background field configurations are: 

\bga
ds^2 &=& \,-\left(1-{M\over r}\right )\,dt^2 +
\left (1-{M\over r}\right )^{-1}\, {K_{SL}dr^2\over 8 r^2} +
{K_{SU}\over4}\left (\,d\theta^2 + 
\sin^2\theta\,d\phi^2\, \right ), \nnu\\ \nnu\\
{\cal A}^{(1)\,\zeta}_\mu &=&\pmatrix {0, &0,&0,& m\,\cos\theta},\>\>\>\>
{\cal A}^{(2)}_{\mu,\,\zeta}=\pmatrix {0,&0,&0,& \pm {n\over
4} \cos\theta },\,\,\,{\cal A}^{(3)\ I}_{\mu}= 0,\nnu\\ \nnu\\
G_{\zeta\zeta}&=&{n\over4\,m},\;\;\;\;\; 
{\Phi}= - \ln r + const.,\>\>\>\>\>\>\> H_{\mu\nu\lambda}=0 .
\label{wz1}
\ega
This background configuration (\ref{wz1}) describes a four dimensional
magnetically charged static black hole solution. In the
asymptotic limit, i.e. $r\to\infty$, the topology of the
spacetime is $R^1\times R^1\times S^2$. 

These solutions (\ref{22}) and 
(\ref{wz1}) satisfy the background field equations of four dimensional
string effective action with nonvanishing cosmological constant
terms,
${2/ K_{SU}} - {2/ K_{SL}}\>\>\>\>
{\rm and}\>\>\>\>\>\>
{2/ K_{SU}}-{4/ K_{SL}}$ respectively,
which indeed break S-duality symmetry of the equations of
motion. The scalar curvatures corresponding to
these background configurations are
respectively given by 
\bga 
{8 \over K_{SU}} - {2 \over K_{SL}} - {3 J^2 \over 2r^4}\>\>\>\>
{\rm and }\>\>\>\>\>\>
{8\over K_{SU}} + {8\, M \over K_{SL}}\, {1\over r}\,\,.\nnu
\ega

In summarizing, we have explored the consequences of
S-duality transformations on the equations of motion with nonzero cosmological
constant. First, we studied a four dimensional action in a
general frame-work. The reduced action (1) could have been
obtained from toroidal compactification of a heterotic string
effective action in higher dimensions. Although these actions do
not necessarily represent supersymmetric theories, 
S-duality invariance would have implied the absence of
cosmological constant. We note that the cosmological
constant term breaks S-duality for the exact conformal field
theory backgrounds and is also responsible for nonvanishing of
$R$ for asymptotically large $r$. 

In this context, let us briefly discuss  the presence of higher order
terms in $\alpha'$ and the consequences of the S-duality transformations
in the equations of motion. 
We write down the  next higher  order term in  $\alpha^{\prime}$ \cite{cal}
to the low energy string effective action (4) as 
\begin{equation}
S'=\int d^4 x \sqrt{-g}\, \eta\, \left(
R_{\mu\nu\lambda\rho}R^{\mu\nu\lambda\rho}\right).
\label{20bb}
\end{equation}
In presence of the higher order term the equation of motion (7) 
gets an additional
contribution 
$${i\over4}  R_{\mu\nu\lambda\rho}R^{\mu\nu\lambda\rho} $$
and similarly eq.(8) is modified with the extra term 
$$\eta\,[ G_{\mu\nu}  +
g_{\mu\nu}\,\bigtriangledown_\alpha\bigtriangledown^\alpha R ], $$
where
$$G_{\mu\nu} = -{1\over2} g_{\mu\nu} R_{\alpha\beta\lambda\rho}
R^{\alpha\beta\lambda\rho} +
2\,R_{\mu\alpha\beta\gamma}R_\nu^{\,\,\alpha\beta\gamma}
-4\,\bigtriangledown_\alpha\bigtriangledown^\alpha R_{\mu\nu} +
2\,\bigtriangledown_\mu\bigtriangledown_\nu R - 4\,
R_{\mu\alpha}R_\nu^{\,\,\alpha} + 
4\,R_{\mu\alpha\nu\beta}R^{\alpha\beta}.$$

We have checked that under the S-duality transformation
(\ref{20a}), eq.(7) with the additional term also breaks
S-duality invariance. The graviton equation along with
the higher order correction term as mentioned above is also not
invariant
under the S-duality. Thus it can be 
argued that the presence of the higher order terms do not 
restore the S-duality invariance in the equations of motion.
Notice that when we dimensionally reduce the terms involving
quadratic in curvature, there will be additional terms in
(\ref{20bb}) involving moduli and gauge fields ( arising from
dimensional reduction ). We have seen that the contribution of
(\ref{20bb}) to equations of motion already breaks the S-duality.
Therefore, even if we explicitly take into account the
contribution coming from moduli and extra gauge fields in the
corresponding equations of motion, the S-duality invariance will
not be restored.

\par In this optics, we propose that vanishing of
$\Lambda$ is closely related with the S-duality symmetry of
string theory if we adopt the ``naturalness'' hypothesis. 
In this context, $\Lambda$
plays a dual role. From the macroscopic point of view, its
smallness in cosmological observations is intimately related to
the fact that our Universe is big and quite flat and in that
sense it tells us about the physics at very large length scale.
On the other hand, when we envisage it from a microscopic point
of view, $\Lambda$ plays the role of a coupling constant in the
string action and it is expected that quantum gravity 
considerations will provide us with an answer to its vanishing
value. In the past, there have been several attempts \cite{4,12}
to explain the smallness of $\Lambda$. Notable among them, the proposal of
Coleman\cite{13} in sequel to the works of Hawking \cite{14} and
Baum \cite{15} has been the focus of attention. However, the
mechanism proposed by Coleman to show 
how the cosmological constant vanishes has received some
criticism \cite{16}. If we construct theories
which respect supersymmetry ( SUSY ) and/or supergravity ( SUGRA
), then $\Lambda$ must be zero; however, these are not exact symmetries
and thus are unable to explain why $\Lambda$ is so small
\cite{4}.  We are aware that models with
SUSY, require the SUSY breaking scale to be of order TeV and
there will be contributions to $\Lambda$ from this breaking. Our
proposal for resolution of the cosmological constant problem is
based on two important hypothesis, $i.e.$ S-duality is an exact
symmetry of string theory and the ``naturalness''.  We recall  that the
cosmological constant, although vanishingly small, does not
satisfy the ``naturalness'' criterion when treated in the framework
of Einstein gravity. However, in the setting of string theory,
we find it exceedingly attractive that $\Lambda $ does fulfill
the criterion of naturalness when we incorporate
S-duality as a symmetry. S-duality as an exact symmetry of
string theory might have far reaching consequences than that
meets the eye. 

\vspace{.2in}
\acknowledgements{ 
Author, S.K. would like to thank Institute of Physics, where most of the work
was done. One of us (J.M.) is grateful to John Schwarz for sharing his 
insights on S-duality and for valuable correspondence. Work of
J.M. is supported by Jawaharlal Nehru Fellowship from Jawaharlal
Nehru Memorial Fund. }


\begin{references}

\bibitem{1} A. Font, L. Ibanez, D. Lust and F. Quevedo, Phys.
Lett. {\bf B249}, 35 (1990); S.J. Rey, Phys. Rev. {\bf D43}, 526
(1991); S. Kalara and D. Nanopoulos, Phys. Lett {\bf B 267}, 343
(1991). A. Shapere, S. Trivedi and F. Wilczek, Mod. Phys. Lett.
{\bf A6}, 2677 (1991); A. Sen, Nucl. Phys. {\bf B404}, 109 (1993), Phys.
Lett. {\bf B303}, 22 (1993), Mod. Phys. Lett. {\bf A8}, 2023
(1993); J. Schwarz and A. Sen, {\bf B312}, 105 (1993), Nucl.
Phys. {\bf B411}, 35 (1994); P. Binetruy, Phys. Lett. {\bf
B315}, 80 (1993); A. Sen, Int. J. Mod. Phys. {\bf A9}, 3707
(1994), Phys. Lett. {\bf B329}, 217 (1994); J. Gauntlett and J.
Harvey, preprint EFI-94-36; L. Girardello, A. Giveon, M. Porrati
and A. Zaffaroni, preprint NYU-TH-94/06/02, NYU-TH-94/12/01;
M. Duff and R. Khuri, Nucl. Phys. {\bf B411}, 473 (1994); J.
Schwarz, preprint CALT-68-1965; J. Harvey, G. Moore and A.
Strominger, preprint EFI-95-01; A. Giveon, M. Porrati and E.
Ravinovici Phys. Rep.{\bf 244C}, 77(1994); J. H. Schwarz,
Proceedings of Workshop on Physics at the Planck Scale, Puri,
December 1994; E. Witten, preprint IASSNS-HEP-95-18, hep-th/9503124. 

\bibitem{2} C. Vafa and E. Witten, preprint HUTP-94-A017; 
N. Seiberg and E. Witten, Nucl. Phys. {\bf B426}, 19 (1994); 
Nucl. Phys. {\bf B431}, 484 (1994); A. Ceresole, R.
D'Auria and S. Ferrara, Phys. Lett. {\bf B339}, 71 (1994);
M. Bershadsky, A. Johansen, V. Sadov and C. Vafa,
preprint HUTP-95-A004; Earlier works in duality includes:
C. Montonen and D. Olive, Phys. Lett. {\bf B72}, 117 (1977); P.
Goddard, J. Nyuts and D. Olive, Nucl. Phys. {\bf B125}, 1
(1977); H. Osborn, Phys. Lett. {\bf B83}, 321 (1979); 
E. Witten and D.Olive, Phys. Lett. {\bf B78}, 97 (1978).

\bibitem{3} G. 't Hooft, {\it Under the Spell of Gauge Principle}
(World Scintific, Singapore, p352, 1994).

\bibitem{4} S. Weinberg, Rev. Mod. Phys. {\bf 61}, 1 (1989), this
article provides beautiful exposition of cosmological constant
problem.  

\bibitem{5} G. Moore, Nucl. Phys. {\bf B293}, 139 (1987); 
E. Alvarez and M.A.R. Osorio, Z. Phys. {\bf C44}, 89 (1989);
E. Alvarez, L. Alvarez-Gaume and Y. Lozano, CERN-Th-7486/94,
hep-th/9410237. 

\bibitem{6} E. Witten, IAS preprint IASSNS-HEP-95-51,
hep-th/9506101. 

\bibitem{7} K. Becker, M. Becker, and A. Strominger,
NSF-ITP-95-07,  hep-th/9502107.

\bibitem{8} J. Maharana and J. Schwarz, Nucl. Phys. {\bf B390},
4 (1993); S. Hassan and A. Sen, Nucl. Phys. {\bf B375}, 103 (1993).

\bibitem{ms} J. Maharana and H. Singh, Phys. Lett. B ( in press
), hep-th/9506213; for related work see, I.
Pinkstone, DAMTP R96-62, hep-th/9505147. 

\bibitem{9} D. Lowe and A. Strominger, Phys. Rev. Lett. {\bf
73}, 1468 (1994); H. Horowitz and D. Welch, Phys. Rev. Lett.
{\bf 71}, 328 (1993); N. Kaloper, Phys. Rev. {\bf D48}, 2598
(1993); for earliar work see; M. Banados, C. Teitelboim and J.
Zanelli, Phys. Rev. Lett. {\bf 69}, 1849 (1992). 

\bibitem{br} B. Bertotti, Phys. Rev. {\bf 116}, 1331 (1959); I.
Robinson, Bull. Acad. Polon. Sci. {\bf 7}, 351 (1959).

\bibitem{10} R. Kallosh, A. Linde, P. Ortin, A. Peet and A. Van
Proeyen, Phys. Rev. {\bf D46}, 5278 (1992); R. Kallosh and A.
Peet, Phys. Rev. {\bf D46}, 5223 (1992).

\bibitem{11} E. Witten, Phys. Rev. {\bf D44}, 314 (1991).
\bibitem{cal} C. G. Callan, D. Friedan, E. Martinec and M.
Perry, Nucl. Phys. {\bf B262}, 593 (1985); A. Giveon, E.
Rabinovici and A. Tseytlin, Nucl. Phys. {\bf B409}, 339 (1993).

\bibitem{12} For detailed references and discussions see, {\it
Euclidean Quantum Gravity}, Eds. G. W. Gibbons and S. W. Hawking (World
Scientific, 1993).

\bibitem{13} S. Coleman, Nucl. Phys. {\bf B310}, 643 (1988); {\bf
B 307}, 854 (1988); J. Preskill, Nucl. Phys. {\bf B323}, 141 (1989);
B. Grinstein, Nucl. Phys. {\bf B321}, 439 (1989).

\bibitem{14} S. W. Hawking, Phys. Lett. {\bf B134}, 403 (1984).

\bibitem{15} E. Baum, Phys. Lett. {\bf B133}, 885 (1983).
\nopagebreak
\bibitem{16} For review see J. Maharana, `` Wormholes and
cosmological constant'' in proceedings {\it Summer School in
High Energy Physics and Cosmology}, Eds. J. C. Pati, S.
Randjbar-Daemi, E. Sezgin and Q. Shafi (World Scientific,
Singapore, 1991).  

\end{references}
\end{document}